\documentclass{ws-procs9x6-cpt16}

\begin{document}  

\newcommand{\refeq}[1]{(\ref{#1})}
\def\etal {{\it et al.}}

\title{Modeling and Analysis of the\\ APOLLO Lunar Laser Ranging Data}

\author{R.D.\ Reasenberg,$^1$ J.F.\ Chandler,$^2$ N.R.\ Colmenares,$^1$\\
N.H.\ Johnson,$^1$ T.W.\ Murphy,$^1$ and I.I.\ Shapiro,$^2$ }

\address{$^1$CASS, University of California, San Diego, La Jolla, CA 92093, USA}

\address{$^2$Harvard-Smithsonian Center for Astrophysics,
Cambridge, MA 02138, USA}

\begin{abstract}
The Earth-Moon-Sun system has traditionally provided the best laboratory for testing the strong equivalence principle.  For a decade, the Apache Point Observatory Lunar Laser-ranging Operation (APOLLO) has been producing the world's best lunar laser ranging data.  At present, a single observing session of about an hour yields a distance measurement with uncertainty of about 2~mm, an order of magnitude advance over the best pre-APOLLO lunar laser ranging data.  However, these superb data have not yet yielded scientific results commensurate with their accuracy, number, and temporal distribution.  There are two reasons for this.  First, even in the relatively clean environment of the Earth-Moon system, a large number of effects modify the measured distance importantly and thus need to be included in the analysis model.  The second reason is more complicated.  The traditional problem with the analysis of solar-system metric data is that the physical model must be truncated to avoid extra parameters that would increase the condition number of the estimator.  Even in a typical APOLLO analysis that does not include parameters of gravity physics,
the condition number is very high: $8 \times 10^{10}$.   
\end{abstract}

\bodymatter

\section{Introduction}

 For over a half century, the Planetary Ephemeris Program (PEP) has been used to analyze solar-system metric data (angles, velocities, distances, etc.).\footnote{We know of no other open-source program capable of such analysis.}  With the availability of data from the Apache Point Observatory Lunar Laser-ranging Operation (APOLLO)\cite{Tom-1}  have come both new opportunities and new challenges.  
The former include the possibility of better estimates for such quantities as $ \dot{G}/G$.  The latter are discussed below.  
 
\begin{figure}[htbp]
\begin{minipage}[hbt]{0.35\textwidth}
\includegraphics[width=1\textwidth]{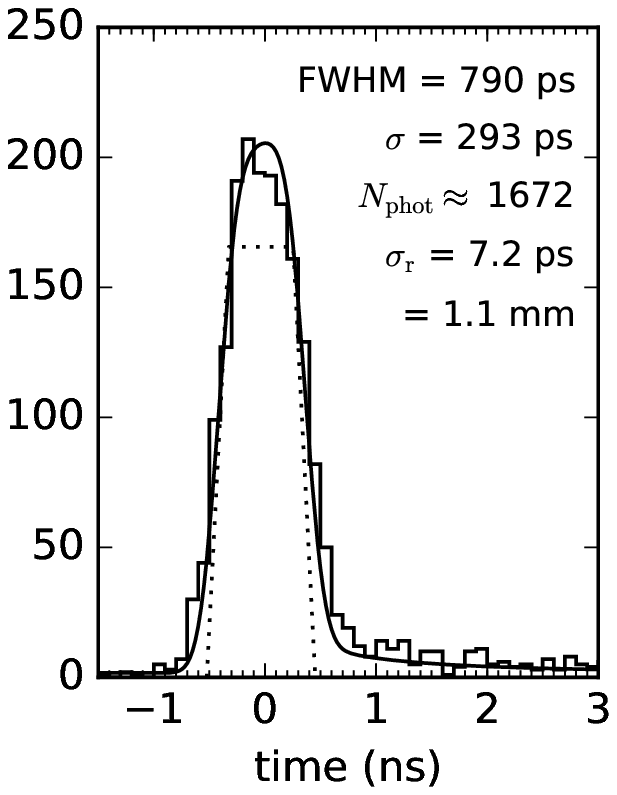}
\end{minipage}
\hfill
\begin{minipage}[hbt]{0.6\textwidth}
\includegraphics[width=1\textwidth]{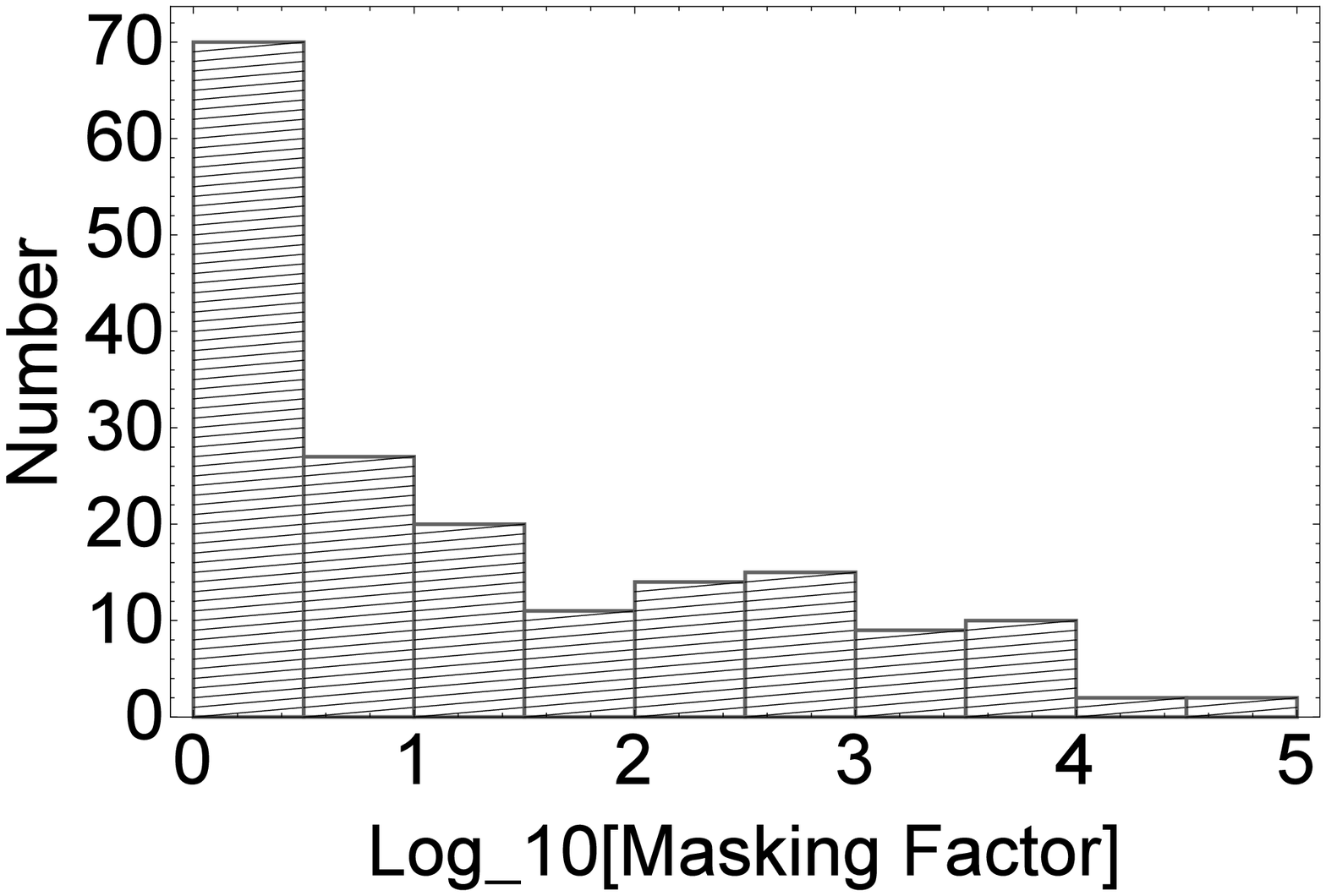}
\end{minipage}
\caption{
Left: return pulse histogram, model (smooth), and reflector trapezoid (dotted).
Right: distribution of masking factors.
}
\label{rdr:fig1}
\end{figure}
 
APOLLO has many features that make possible its enhanced precision including a large telescope ($\phi$3.5 m) for collecting returned photons, a short (100 ps) transmitted pulse length, and a detector comprising a 4 $\times$ 4 array of avalanche photodiodes (APD) that are efficient, allow the detection of more than one photon per pulse, and provide telescope pointing feedback.
 
The finite size of the retroreflector arrays on the Moon combined with the deviation of the line of sight from the normal to the targeted array results in a “return light pulse” that in some cases is considerably broader than the transmitted pulse (see Fig.~\ref{rdr:fig1} left). In the first data reduction step, the photon return times from a single observing span yield a normal point (NP).  Each photon round-trip time is reduced to a residual with a working ephemeris.  A model of the returned pulse is then fit to the collection of residuals.  That model includes a single delay parameter and a template of the broadening due to the known observing geometry convolved with the shape of a typical transmitted pulse.  The estimated delay parameter is added to the nominal round-trip light time from the same ephemeris at an epoch near the middle of the data-taking span to form the NP.  Thus, the working ephemeris need not be highly accurate.

The sub-cm uncertainty of the APOLLO NP is made up of three components.  First is the statistical error from the above fitting of the return-pulse model to the timing residuals.  The second comes from the uncorrelated part of the fluctuations in the estimated delay correction among the 16 APD detectors.  This is likely due to the response of the APD-specific electronics implementation.  The third is the common fluctuations in the response of the electronics. This might come, for example, from jitter in the timing clock.  Components 1 and 2 are measured and component 3 must be inferred from the post-fit residuals or from an absolute calibration subsystem such as the one that is currently being added to the APOLLO measuring system.

\section{Data modeling and analysis}

The APOLLO NPs have a fractional error of $ \sigma (L)/L\approx 10^{-11}$.  Even in the relatively clean Earth-Moon system, there are many ``effects'' at this level. However, as is often the case with solar-system metric data, some desired model extensions cannot be made because they bring with them the need to estimate additional parameters.  When included in the estimator, these parameters so increase the condition number (see below) that they are not practical to use.  Our analysis includes both lunar and non-lunar data and model components (e.g., masses and orbital elements of bodies) and is based on weighted least squares (WLS) fitting.  

The current baseline analysis uses all of the available data to fit 180 parameters, but that set excludes the parameters of interest for gravity science $(\eta_{SEP},~ \beta ,~  \dot{G}$, SME~coefficients, etc.), pending all else being thoroughly checked. From this analysis, four warning 
signs emerge.  The ``condition number,'' $N_C$, the ratio of the largest to the smallest eigenvalue of the normalized WLS coefficient matrix, is $8 \times 10^{10}$.  
Thus, in a numerical operation like solving the normal equations, the effect of numerical noise may be magnified by  $10^{11}$.  Since we currently run PEP on a machine that 
uses the IEEE 80-bit format for floating point numbers (19 digits), the high value of $N_C$ is not preventing the analysis from going forward.  The second sign is that there are some very 
large ``masking factors,'' defined for each estimated parameter as $\mathfrak{M}_k= \sigma _k  ({\rm full~analysis})  / \sigma _k ({\rm single~parameter~estimate})$.   Figure~\ref{rdr:fig1} (right) shows the distribution of $\mathfrak{M}$, which implies that the sensitivity-matrix elements must be unusually  accurate, reaching one part in 10$^7$ for a few.  Both $N_C$ and $\mathfrak{M}$ manifest the cumulative effect of parameter correlations.

Third, the iterated WLS estimator should converge very quickly in the linear regime; currently there are only small adjustments needed. In fact, it does not, and we have run PEP for several tens of iterations in an attempt to reveal a pattern.  This slow convergence suggests that some sensitivity matrix elements (often called the partial derivatives)  are not sufficiently accurate (see below).  Fourth, the post-fit residuals are highly systematic, which suggests the need to enhance the PEP model.

For some of the estimated parameters (e.g., lunar eccentricity), the sensitivity matrix includes components found by numerically integrating the variational equations in parallel with the equations of motion.  We have added additional small ``indirect terms'' to the integrand and shown by numerical methods that the integrated variational equations no longer should be suspected of being the problem. While this has resulted in better convergence, it is still slower than expected. Until these four issues are resolved, there is no reason to attempt to estimate the parameters of gravity physics.

\section{Planned improvements}

There are several known small contributions to the model of the LLR observable that have been added recently or are planned for the near future.  Among those pending are: (1) rotational deformation due to the time-varying centrifugal potential from the motion of the Earth's rotation pole around its mean location (radial, horizontal: (2.5, 0.7 cm); 2) fluid loading of the crust by atmospheric pressure and groundwater (radial, about 1 cm); (3) shift of the center of mass of Earth with respect to the center of figure, as defined by the ensemble of tracking stations, due to the seasonal deposit of frozen water (N-S, about 1 cm); and (4) the effect of radiation pressure on the lunar motion ($0.36 \cos(D)$ cm in the Earth-Moon separation, where $D$ is the Moon-Earth-Sun angle).

Our strategy for addressing the first three of these phenomena includes validation using continuous measurements of displacement by a nearby GPS station (P027,  2.5 km away) and of local gravity by the Apache Point superconducting gravimeter.  Although neither of these instruments gives a direct measure of the displacement of the Apache Point Telescope, models of their response to the three drivers will test the corresponding models in PEP.  Also pending is a merge of the lunar integration (orbit and rotation) with the integration of the planets, including the Earth-Moon system.  This is to replace an inconvenient iterative scheme that is believed to be adequately accurate for the analysis of the current data.

 \section*{Acknowledgments}
 This work was supported in part by NASA grant NNX12AE96G and NSF grant PHY1068879.


\begin{thebibliography}{x}

\bibitem{Tom-1}
T.W.\ Murphy, 
Rep.\ Prog.\ Phys.\ {\bf 76}, 076901 (2013).

\end{thebibliography}
\end{document}